\documentclass[conference]{IEEEtran}
\usepackage{graphicx}
\usepackage{newtxmath}
\usepackage{url}
\usepackage[nobreak]{cite}
\usepackage{xspace}

\def\ChangeBeadsThreader{ChangeBeadsThreader\xspace}
\def\CBT{CBT\xspace}
\newcommand{\etal}{\textit{et al.}\xspace}
\newcommand{\maru}[1]{\raise0.2ex\hbox{\textcircled{\scriptsize{#1}}}}
\newcommand{\M}[1]{\textsf{#1}}

\begin{document}
\title{ChangeBeadsThreader: An Interactive Environment\\for Tailoring Automatically Untangled Changes}

\author{\IEEEauthorblockN{Satoshi Yamashita, Shinpei Hayashi, and Motoshi Saeki}
\IEEEauthorblockA{School of Computing, Tokyo Institute of Technology, Tokyo 152--8550, Japan\\
Email: \{yamashita, hayashi, saeki\}@se.cs.titech.ac.jp}}

\maketitle
\begin{abstract}
To improve the usability of a revision history, change untangling, which reconstructs the history to ensure that changes in each commit belong to one intentional task, is important.
Although there are several untangling approaches based on the clustering of fine-grained editing operations of source code, they often produce unsuitable result for a developer, and manual tailoring of the result is necessary. 
In this paper, we propose \ChangeBeadsThreader~(\CBT), an interactive environment for splitting and merging change clusters to support the manual tailoring of untangled changes. 
\CBT provides two features: 
1) a two-dimensional space where fine-grained change history is visualized to help users find the clusters to be merged and 
2) an augmented diff view that enables users to confirm the consistency of the changes in a specific cluster for finding those to be split. 
These features allow users to easily tailor automatically untangled changes.
\end{abstract}
\begin{IEEEkeywords}
    change untangling; version control systems;
\end{IEEEkeywords}

\IEEEpeerreviewmaketitle

\section{Introduction}\label{s:introduction}
Version control systems such as Git \cite{Git} are widely used in software development.
Developers often mix changes in different intentional tasks in one commit, which results in a \emph{tangled commit}\cite{herzig2013}.
Existing studies pointed out that tangled commits render the reuse, revert, and understanding of changes complex \cite{hayashi-iwpse-evol2010} and lead to difficulties in code review \cite{bacchelli2013}, software evolution \cite{berczuk}, and mining software repositories \cite{Nguyen2013}.

To prevent tangled commits, \emph{change untangling}, i.e., decomposing tangled commits in a repository to ensure that each commit consists of changes of one intentional task, is important \cite{herzig2013}.
Existing change untangling approaches are classified into two types:
  1) given a tangled commit, finding boundaries among the changes in the commit and separating them \cite{Nguyen2013,kirinuki-jit-untangling,barnett2015,kirinuki-past-changes,Kreutzer-clustering,coen2018,sothornprapakorn2018}, and
  2) given a sequence of finer-grained (edit-level) changes, clustering them and forming untangled commits based on the clusters \cite{Dias2015,jmatsu-iwpse2015}.
Although the latter approach requires a sequence of finer-grained changes rather than commits, 
it is efficient because it can capture the temporal characteristic of changes \cite{Dias2015}.

A challenge in automated change untangling is that providing \emph{perfect} results is difficult.
This is because the concept of \emph{tasks}, which acts as a key role in commit construction, is ambiguous.
Therefore, a manual tailoring process by developers is necessary to fix the result of an automated change untangling technique.
Although Dias \etal provided a simple tool for fixing change untangling results \cite{Dias2015}, the tool did not focus on its usability, and the examinee pointed out issues due to the information overload when a large number of changes are given.

In this paper, we propose a tool named \emph{\ChangeBeadsThreader}~(\CBT), an interactive environment for splitting and merging change clusters to support the manual tailoring of untangled changes. 
\CBT has two features: 
1) a two-dimensional space where a fine-grained change history is visualized and 
2) an augmented diff view that enables users to confirm the consistency of the changes in the specified clusters.
These two features respectively help users find the set of clusters to be merged and changes to be split in a specific cluster, and they enable efficient tailoring of untangled changes.

The remainder of this paper is organized as follows.
Section~\ref{s:relatedworks} discusses related work.
Section~\ref{s:require} explains the obstacles in tailoring untangled changes and defines the requirements for \CBT.
Section~\ref{s:ChangeBeadsThreader} describes the detail of \CBT.
Section~\ref{s:conclusion} concludes this paper and summarizes future work.

\section{Related Work}\label{s:relatedworks}

Tangled commits lead to bad effects in software development.
Bacchelli \etal \cite{bacchelli2013} reported that tangled commits decrease the quality and increase the required time in code review.
Herzig \etal \cite{herzig2013} investigated the frequency and impact of tangled changes.
They reported that more than 15\% of bug-fix commits in five Java open source projects were tangled, which causes at least 16.5\% of source files were incorrectly associated with bug reports.

To mitigate these issues, several approaches to untangle changes in a commit (the first type of untangling approaches) have been proposed \cite{Nguyen2013,kirinuki-jit-untangling,barnett2015,kirinuki-past-changes,Kreutzer-clustering,coen2018,sothornprapakorn2018}.
For example, Barnett \etal \cite{barnett2015} proposed a change untangling technique for C\# programs.
Their approach extracts the def-use dependency among methods and utilizes it for detecting related changes.
A tool named ClusterChanges has been implemented to support the code review process for tangled commits.

Some approaches utilize a fine-grained change history for change untangling.
Dias \etal \cite{Dias2015} proposed and implemented an automatic change untangling technique for a fine-grained change history.
They defined 13 metrics on such a history and investigated which metrics contribute to change untangling.
As a result, the metrics of \M{timeDistance}, \M{numberOfEntriesDistance}, and \M{sameClass} were useful.

Several approaches to visualize fine-grained changes have been proposed.
Yoon \etal \cite{yoon2013,yoon2015} proposed a tool named Azurite to support the selective undo operation with visualizing a sequence of fine-grained changes.
Omori and Maruyama \cite{OperationReplayer} proposed OperationReplayer to replay an editing history of source code for understanding software evolution.
Barik \etal \cite{CommitBubble} proposed a concept named Commit Bubbles, which enables developers to manipulate a commit history easily.
However, to our knowledge, no tools have been proposed to support the tailoring process of commits automatically constructed from a fine-grained history.

\section{Problems and Requirements}\label{s:require}
In this paper, we adopt the approach of merging a fine-grained change history obtained from change recording tools such as ChangeMacroRecorder \cite{ChangeMacroRecorder} or Fluorite \cite{Fluorite}.
In this section, we explain the problems and requirements of change untangling for a fine-grained change history.

\subsection{Running Example}\label{ss:example}

\begin{figure}[t]\centering
    \includegraphics[width=8.7cm]{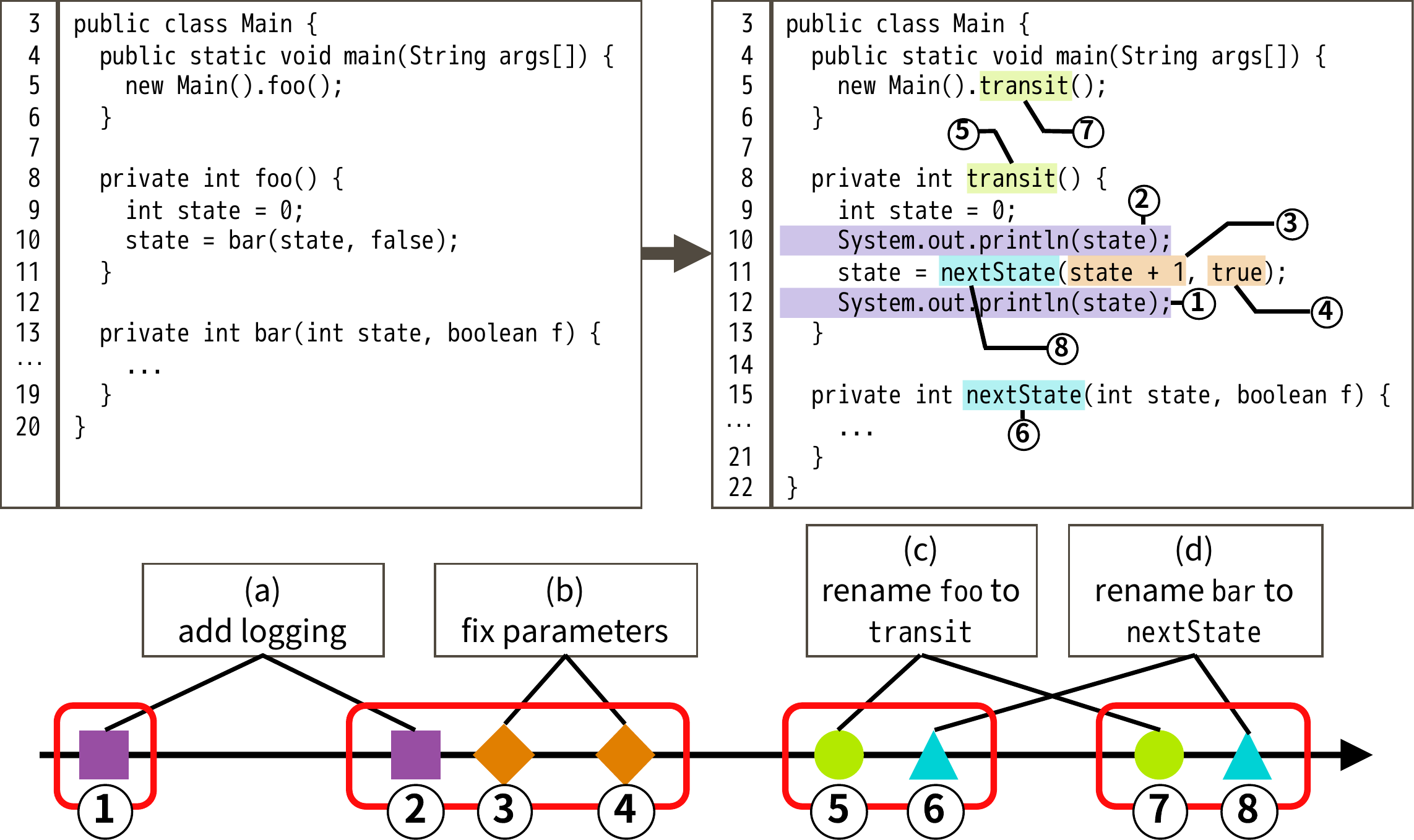}
    \caption{Example of change untangling.}\label{fig:input_exp}
\end{figure}
Figure~\ref{fig:input_exp} shows an example of the change untangling.
The source code before and after applying the changes are respectively shown in the upper left and the upper right of the figure.
The changes (a) \maru{1} and \maru{2} add logging method invocations to show the value of \texttt{state},
            (b) \maru{3} and \maru{4} fix the parameters of a method invocation,
            (c) \maru{5} and \maru{7} rename the method \texttt{foo} to \texttt{transit}, and
            (d) \maru{6} and \maru{8} rename the method \texttt{bar} to \texttt{nextState}.
These changes are also shown in the timeline at the bottom of the figure.

By applying a clustering-based automated change untangling technique which utilizes computed metric values on changes, we can have a separation from the given sequence of changes.
The red rounded rectangles in the lower part of Fig.~\ref{fig:input_exp} show the clusters obtained from such a technique.
In this result, the changes \maru{2}, \maru{3}, and \maru{4} are incorrectly clustered because the executed time between the changes \maru{2} and \maru{3} is close.
The changes \maru{5} and \maru{6}, and the changes \maru{7} and \maru{8} are also incorrectly identified as clusters in the same way.

It is difficult to construct a perfect automated change untangling technique because the concept of \emph{tasks}, which acts as a key role in the commit construction, is ambiguous.
Therefore, a manual tailoring process by developers is necessary to fix the result of an automated change untangling technique.

\subsection{Challenges}

To fix the result, we need to tackle two problems.
The first one is how to split clusters.
In the tailoring process, we need to identify the changes to be excluded from a tangled cluster, e.g., the one consisting of \maru{2}, \maru{3}, and \maru{4} in the previous example.
One typical way to find such changes is to check the \emph{diff} of all the changes in the cluster and find anomalies.
In the case above, developers can identify that the changed line that adds a method invocation printing \texttt{state}~(Line 10) should be excluded and be merged into another cluster.
However, determining the changes to be excluded by associating the lines with the changes is a time-consuming and troublesome process, and developers may naively end up checking the diff of the three changes one by one.

The second problem is how to merge clusters.
We need to merge two clusters if we find the changes in them belong to the same task; e.g., the new cluster consisting of \maru{2} obtained from the split above should be merged into the cluster consisting of \maru{1}.
However, finding a cluster where the target cluster should be merged is also time-consuming because all the clusters except the target cluster can be the candidates, and developers need to check the changes in each cluster one by one.
For example, developers need to discover the cluster for the changes to add invocations to print \texttt{state}~(Line 10), i.e., the one consisting of \maru{1} by confirming the changes of each cluster. 
This process becomes hard if the number of changes becomes large.

To sum up, the following problems should be addressed in tailoring the automated results:
   1) checking the changes one by one in the target cluster to find the ones to be excluded and
   2) checking the clusters one by one to find the one that the target cluster should be merged.

\subsection{Requirements}\label{ss:require}

To solve these problems, desired tools need to satisfy the following requirements:
    1) for splitting clusters, users can easily recognize anomaly changes in the candidate clusters to be split, and
    2) for merging clusters, users can easily recognize changes close to each other across clusters.

\begin{figure*}[t]\centering
    \includegraphics[width=13cm]{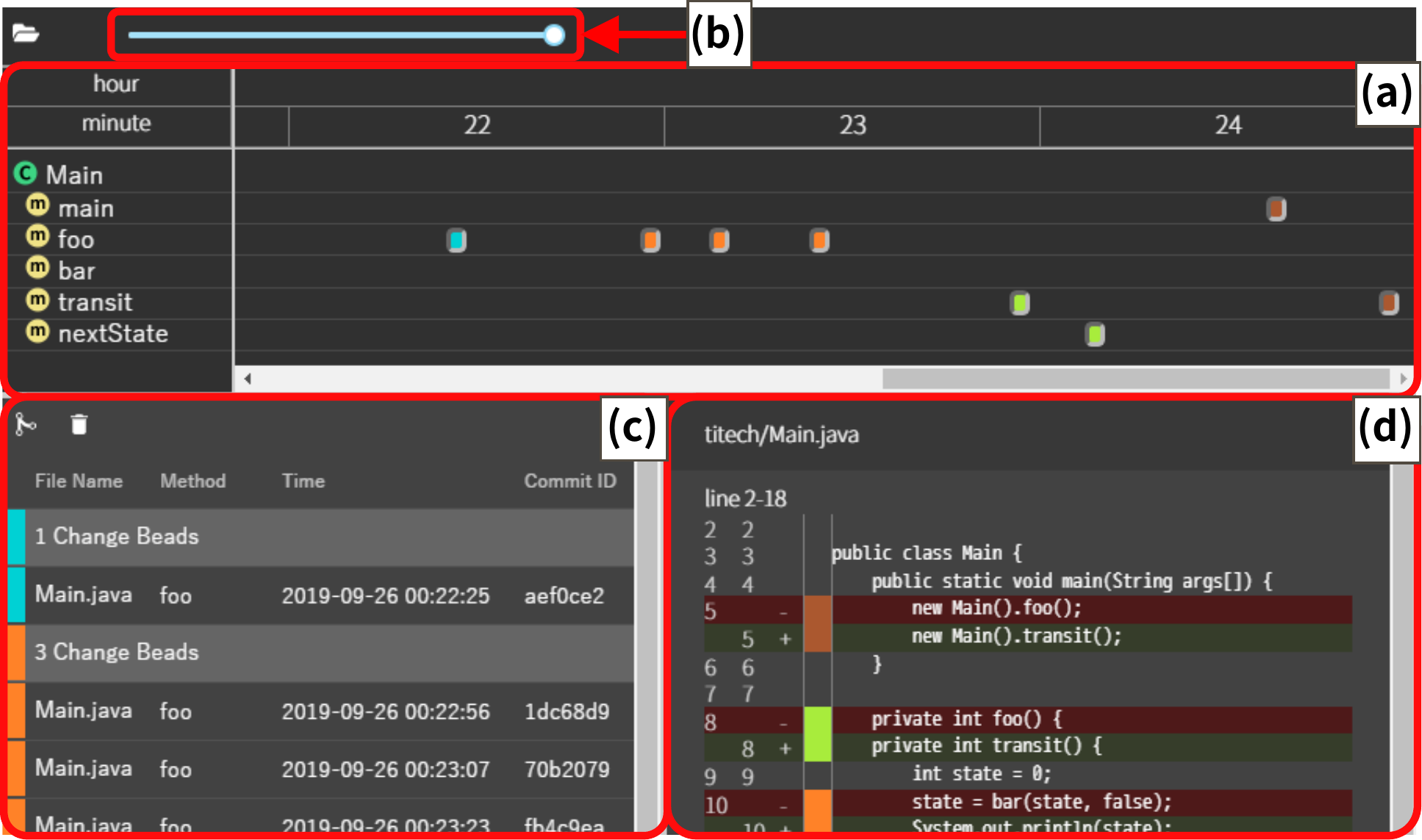}
    \caption{Screenshot of \ChangeBeadsThreader. (a) Map pane. (b) Slider to manipulate the scale. (c) List pane. (d) Diff pane.}\label{fig:ChangeBeadsThreader}
\end{figure*}

\section{\ChangeBeadsThreader}\label{s:ChangeBeadsThreader}

\subsection{Overview}\label{ss:overview}

We have developed a tool named \ChangeBeadsThreader~(\CBT), which is an interactive environment for supporting the manual tailoring of automatically untangled changes\footnote{Available at \url{https://github.com/salab/ChangeBeadsThreader}}.
In developing \CBT, we conceptualized fine-grained changes and their clusters as beads and beadworks, respectively.
The commit construction process is then regarded as beads threading.
Hereafter, we call fine-grained changes treated in \CBT \emph{change beads}.
In \CBT, each change bead belongs to a specific cluster.
Each cluster has its own color, which is automatically assigned by the tool.
Change beads that belong to a cluster are rendered with the color of the cluster.

Figure~\ref{fig:ChangeBeadsThreader} shows a screenshot of \CBT, which consists of three panes.
The {\it Map} pane~(\maru{1}) visualizes the change beads in the given history on the two-dimensional plane to assist the merging of change clusters.
The scale of the timeline on the horizontal axis can be manipulated by operating the slider~(\maru{2}).
The {\it List} and {\it Diff} panes are for showing the information of the change beads of the selected clusters in Map pane.
The List pane~(\maru{3}) lists the changes and shows their details such as the belonging cluster, the belonging class and method, and the time when it was committed.
The Diff pane~(\maru{4}) shows an augmented {\it diff} of the changes, which allows users to check the consistency of the selected clusters.

\begin{figure}[t]\centering
    \includegraphics[width=4.5cm]{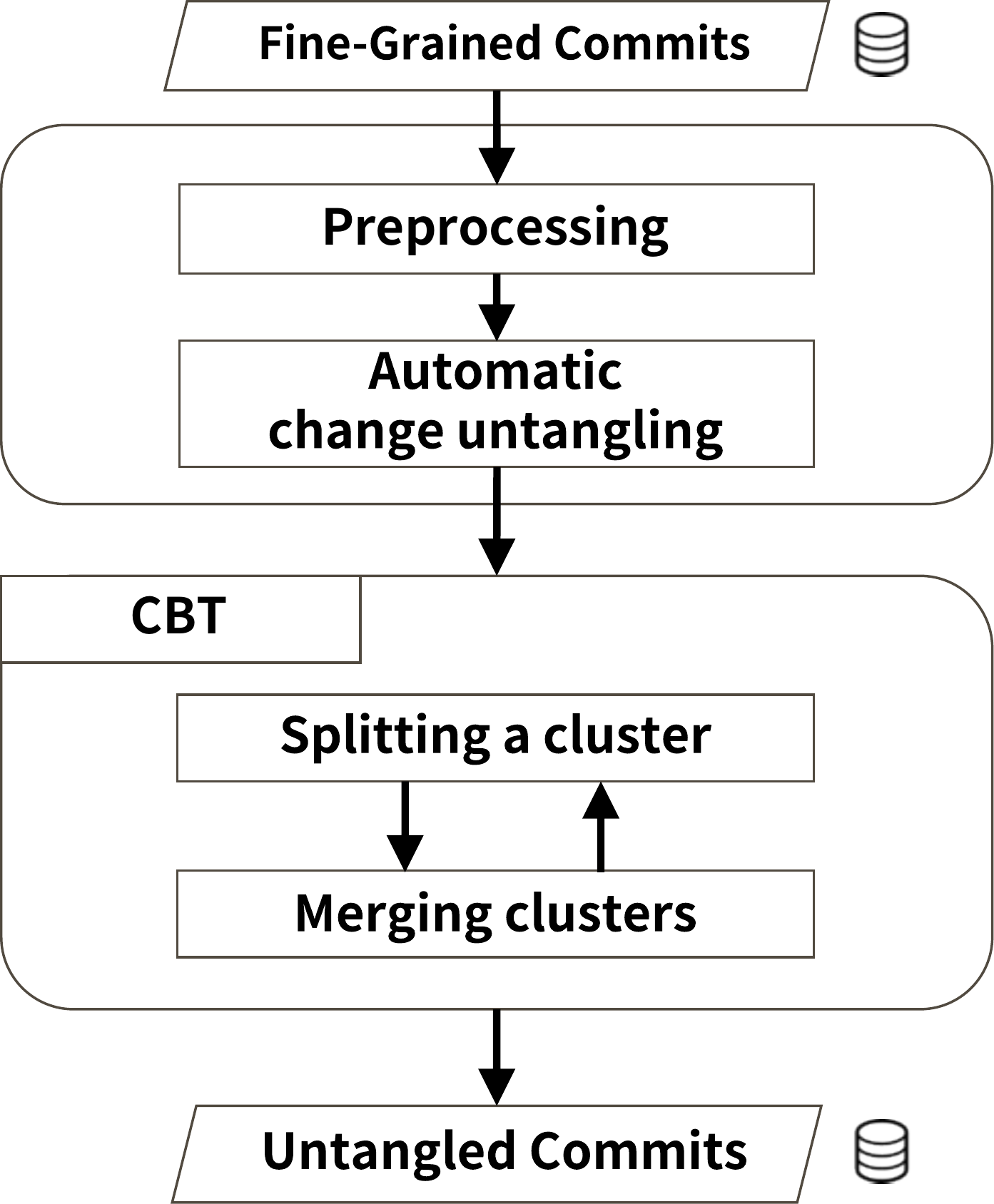}
    \caption{Flowchart of \ChangeBeadsThreader.}\label{fig:flowchart}
\end{figure}
Figure~\ref{fig:flowchart} shows the usage flow of \CBT.
\CBT takes a Git repository consisting of fine-grained commits of a Java program as its input.
It regards a sequence of commits as a sequence of fine-grained changes that are about to be committed as a single commit.
Its output is another repository consisting of untangled commits, which are created by reordering and squashing the given fine-grained commits based on the idea of the history refactoring \cite{history-refactoring}.

First, \CBT preprocesses the input repository and applies an automated change untangling technique to build initial change clusters.
The user sees the displayed clusters of change beads and finds one that is suspected as tangled.
When selecting a cluster, its {\it diff} representation is displayed.
The user splits it or merges it with another cluster until the user agrees with the result.
Finally, the obtained clusters form untangled commits and stored into the resulting repository as an output.

\subsection{Preprocessing}\label{ss:eachProcess}

As a preprocessing, \CBT identifies some commits containing source files that their parsing fails to be excluded.
They are squashed with their neighbor commits so that the changes made in them are still available in the repository.
This is to avoid failing to identify the classes and methods that the changes in the commit belong to.
After that, \CBT extracts the information of each commit.

\subsection{Automatic Change Untangling}\label{ss:initialClustering}

\begin{figure*}[t]\centering
    \includegraphics[width=18cm]{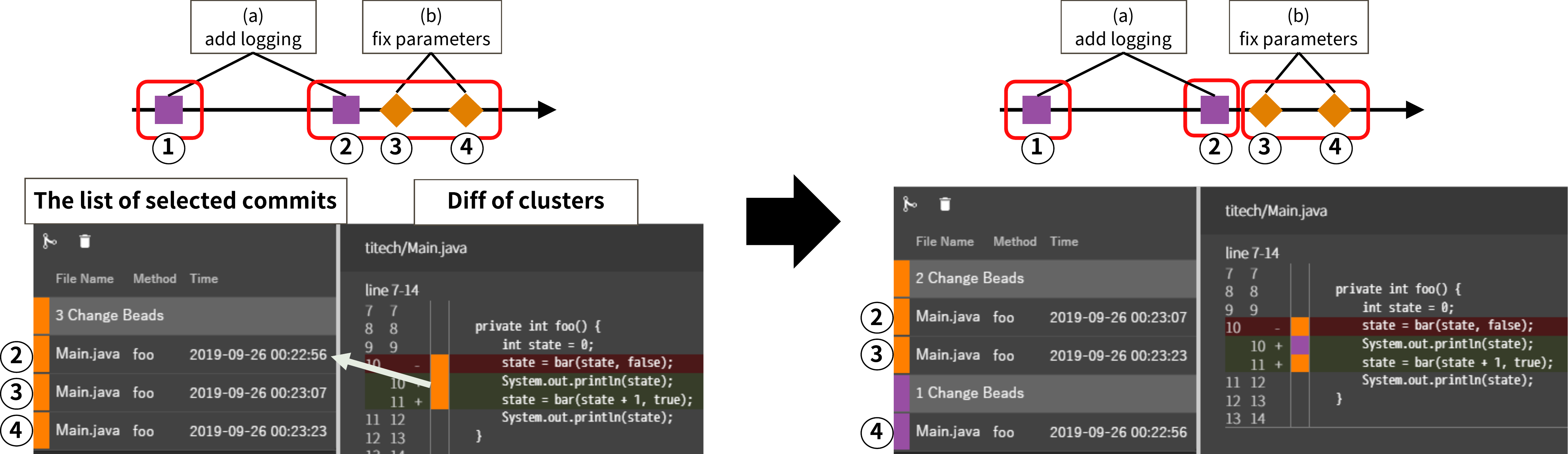}
    \caption{Splitting a cluster.}\label{fig:cluster_split}
\end{figure*}

\begin{table}[t]\centering
    \caption{Metrics for Calculating Distance between Commits}\label{tab:calc_distance}
    {\tabcolsep=5pt\begin{tabular}{lp{4.9cm}} \hline
        Name               & Description \\\hline
        \M{timeDistance}   & Seconds between two given commits \\
        \M{numberOfEntriesDistance} & Number of commits in between two given commits \\
        \M{sameClass}      & Whether the changes in two given commits belong to the same class (0 or 1) \\
        \M{sameMethod}     & Whether the changes in two given commits belong to the same method (0 or 1) \\\hline
    \end{tabular}}
\end{table}
After the preprocessing, initial clusters are constructed by applying an automated change untangling technique.
We followed the study by Dias \etal \cite{Dias2015} on a clustering-based change untangling technique.
They defined several metrics of fine-grained changes and applied a clustering algorithm using the defined metrics.
The initial clustering in \CBT follows a simplified variant of their technique.
We selected a subset of their metrics, as shown in Table~\ref{tab:calc_distance}.
The first two metrics quantify the temporal characteristic of changes whereas the other two metrics do the structural characteristic.

Using the metrics, \CBT measures the distance for all the pairs of commits:
\[
    \mathit{distance}(c_1, c_2) = \sum_{i=1}^{|M|} \, \alpha_i M_i(c_1, c_2)
\]
where $c_1$ and $c_2$ are the given commits, $M$ is the set of metrics, $M_i(c_1, c_2)$ is the value of $i$-th metric for the given commits, and $\alpha_i$ is the weight of $i$-th metric.
If the distance is lower than the threshold, they belong to the same cluster.

\subsection{Splitting a Change Cluster}\label{ss:support_split}

The Diff pane supports the split of clusters.
Figure~\ref{fig:cluster_split} shows an example cluster to be split.
The left side of the figure shows the cluster before splitting, 
which consists of three changes: \maru{2}, \maru{3}, and \maru{4}, and here the user is trying to split it.
All the changes in the cluster are listed in the List pane.
Also, the changes are shown as the unified diff format in the Diff pane.
Rather than a normal diff obtained from Git, the diff shown in the Diff pane is augmented with the clustering information; 
the colors of the belonging clusters are attached to the changed lines.
On the left side of the figure, the orange color is attached to all three changed lines because all three changes belong to the same cluster.

The user found that the first and the last changed lines fix the parameters of the method \texttt{bar} to change the behavior 
whereas the middle changed line adds a new method invocation for a debugging purpose.
From this observation, the user identified that the middle changed line has a different purpose from the other two changed lines and is an anomaly.
Therefore, the user selects this anomalous change so that it is excluded from the cluster and forms a new cluster, as shown in the right side of the figure.

In this way, users can identify change clusters to be split without looking at the diff of each change separately.

\subsection{Merging Change Clusters}\label{ss:support_integration}
\begin{figure*}[t]\centering
    \includegraphics[width=18cm]{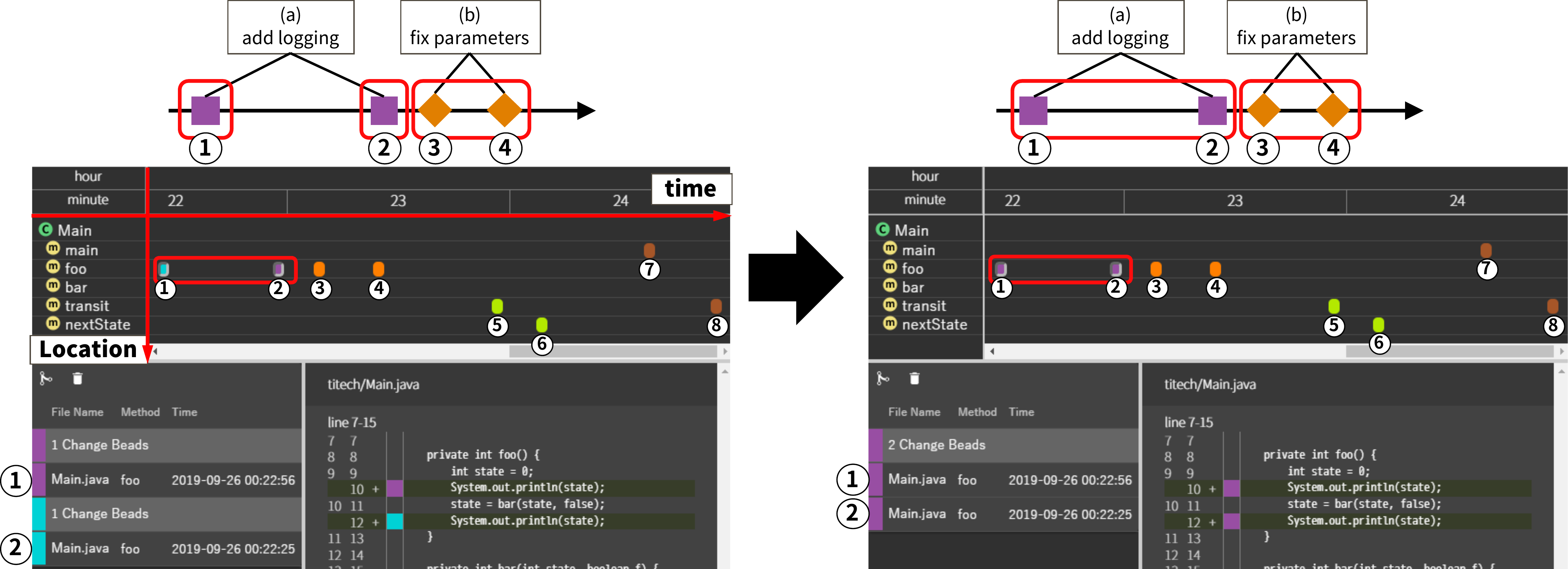}
    \caption{Merging clusters.}\label{fig:cluster_integration}
\end{figure*}

The change beads visualization can be used to see all the changes at a glance and find candidate clusters where a cluster should be merged to.
Figure~\ref{fig:cluster_integration} shows an example of the change beads visualization and how it can be utilized for the cluster merging.
In this pane, the horizontal axis corresponds to the timeline representing \emph{when} each change happened 
whereas the vertical axis corresponds to the locations representing \emph{where} each change was performed.
These two axes are based on the study by Dias \etal \cite{Dias2015}; 
they showed that metrics based on temporal and structural distance are effective for clustering-based change untangling.
We believe that these metrics are also effective in visualizing the distance between change beads so that they are adopted as the axes of our two-dimensional space.
Change beads are displayed in this two-dimensional space in different colors according to their belonging cluster.

On the left side of Fig.~\ref{fig:cluster_integration}, 
the user searches for the candidates, where the cluster consisting of the change \maru{2} should be merged, using the Map pane.
Other than the cluster of \maru{3} and \maru{4}, 
which is the origin of the newly introduced target cluster, the cluster of \maru{1} seems to have a change performed to the same method as \maru{2}, 
and the changes \maru{1} and \maru{2} are relatively closer in both the temporal and structural space; they were performed sequentially at the same method.
As shown in the lower-left side of Fig.~\ref{fig:cluster_integration}, 
the diff of all the changes in these two clusters is checked using the Diff pane as explained above.
Since the user decides that the two clusters are identical,
they are merged as shown in the right side of Fig.~\ref{fig:cluster_integration}.

In this way, users can visually recognize clusters that are close in the temporal and the structural space, 
which helps users to find clusters where the target cluster should be merged to.

\subsection{Implementation}\label{ss:implement}

The implementation of \CBT uses Eclipse Java development tools \cite{JDT} and JGit \cite{JGit} to retrieve change information.
\CBT itself is implemented as a desktop application using the Electron framework \cite{Electron}.

\section{Conclusion}\label{s:conclusion}
In this paper, we propose \ChangeBeadsThreader, a tool to assist developers in tailoring automatic change untangling results.
The tool provides the Diff pane that can confirm the consistency of changes within a specific cluster and the Map pane that visualizes fine-grained change history on a two-dimensional plane of the temporal and structural space.
These panes support the discovery of the clusters to be merged and those to be split, which are the main task in tailoring untangled changes.

Our future work to further improve the usability of \CBT is as follows:
\begin{itemize}
    \item \emph{New view} to visualize an overview of clusters to help users identify where they start to fix.
    To identify the cluster to be fixed first, outlining the abstracts of all clusters at a glance is necessary.
    \item \emph{Highlighting change beads} that are close to those in the selected cluster for supporting more the cluster merging.
    \item \emph{Visualizing metric values}.
    \CBT uses metrics only for extracting temporal and structural characteristics.
    For example, visualizing the def-use dependency of variables by drawing the connection between change beads should be beneficial because it helps users to find candidate change beads that should be merged but located at far.
\end{itemize}

\section*{Acknowledgments}
This work was partly supported by JSPS Grants-in-Aid for Scientific Research Number JP18K11238.

%\bibliographystyle{IEEEtran}
%\bibliography{references}


\end{document}